\documentclass[10pt,preprint]{aastex}

\shorttitle{Relativistic Jet in Wind}
\shortauthors{Hardee \& Hughes}
\slugcomment{To appear in The Astrophysical Journal v.583 January 20}

\begin{document}

\input colordvi.sty 

\baselineskip 10pt
\parskip 2pt

\title{The Effect of External Winds on Relativistic Jets}

\author{Philip E. Hardee}
\affil{Department of Physics \& Astronomy, University of Alabama,
       Tuscaloosa, AL 35487} 
\email{hardee@athena.astr.ua.edu}

\author{Philip A. Hughes}
\affil{Astronomy Department, University of Michigan, Ann Arbor, MI 48109}
\email{hughes@astro.lsa.umich.edu}

\begin{abstract}
\baselineskip 10pt

Relativistic jets in Galactic superluminals and extragalactic AGN may
be surrounded by a wind near to the central engine.  Theoretical
analysis and numerical simulation reveal considerable stabilization  of
relativistic jet flow by a wind to helical and higher order asymmetric
modes of jet distortion.  When velocities are measured in the source
(inlet) frame, reduction in the absolute velocity difference between
jet and wind, $\Delta v = v_{jet} - v_{wnd}$, provides stabilization in
addition to stabilization provided by a high jet Lorentz factor, but a
high Lorentz factor wind is not needed to stabilize a high Lorentz
factor jet.   However, the fundamental pinch mode is not similarly
affected and knots with spacing a few times the jet radius are
anticipated to develop in such flows.  Thus, we identify a mechanism
that can suppress large scale asymmetric structures while allowing
axisymmetric structures to develop. Relativistic jets surrounded by
outflowing winds will be more stable than if surrounded by a stationary
or backflowing external medium.  Knotty structures along a straight jet
like that in 3C\,175 could be triggered by pinching of an initially low
Mach number jet surrounded by a suitable wind.  As the jet enters the
radio lobe, suppression of any surrounding outflow or backflow
associated with the high pressure lobe triggers exponential growth of
helical twisting.

\end{abstract}


\vspace{-1.5cm}
\section{Introduction}

Numerical studies (see \citet{mku01} and references therein) and
theoretical work \citep{bb99} indicate that relativistic jets may be
surrounded by a more slowly moving wind.  Flow in the environment
outside the jet can have important consequences for its stability, and
the twisted normal mode structures that appear on resolved jets.  In
this paper we present results from a numerical simulation of a
precessed elliptically distorted relativistic jet.  The simulation was
designed to explore the interaction between helical twisting arising
from precession and the twin helically twisted filaments predicted to
accompany elliptical distortion.  In the simulation the jet develops a
velocity shear (wind) layer which has significant consequences for the
evolution of the initial perturbation.  The results are relevant to the
development of structures in relativistic Galactic and extragalactic
jets.

\vspace{-0.5cm}
\section{Simulation Setup}

The simulation used a new numerical 3D relativistic hydrodynamical
code. A description of the code along with validation tests and
comparison to previous 2D simulation results to establish the
reliability of the techniques can be found in \citet{hmd02}.  The
present simulation employed a uniform 3D Cartesian grid resolved into
108 $\times $ 108 $\times $ 549 zones. The 108 zones along the
transverse Cartesian axes span a distance of $\sim$~8R, the 549 zones
span a distance of $\sim$~41R along the jet ($z$) axis, and 27 zones
span a jet diameter, 2R.  In the simulation the jet is initialized
across the entire grid, jet and ambient proper densities are equal, the
jet is in pressure balance with the ambient medium, the Lorentz factor
$\gamma$~=~7.47 (v$_{\rm j}$ = 0.991~c), the adiabatic index
$\Gamma$~=~13/9 and the sound speed a~=~0.62~c.  A precessional
perturbation of angular frequency $\omega$R/$v_{\rm j}$~=~0.5 is
applied at the inlet by imposing a transverse component of velocity
$v_{\perp}$~=~0.0025~c.  An elliptical perturbation is also applied at
the inlet by imposing an elliptical cross section of eccentricity 0.51
on the jet at the inlet with rotation of the major axis at the same
angular frequency as the precessional perturbation.

In the simulation outflow boundary conditions are used except on the
plane where the jet enters the computational grid. Inflow is imposed on
the plane $z=0$, and involves cells cut by the jet boundary for which
state variables must be established through a volume-weighted average
of the internal and external values. To avoid a `leakage' of jet
momentum into the ambient material, fixed, initial values are used
across the entire boundary plane at every time step.  However, the high
momentum density of the jet (approximately an order of magnitude higher
than that of the $\gamma=2.5$ flows discussed by \citet{hhrg01}) means
that even modest numerical viscosity is capable of transferring a
significant momentum to the ambient material beyond the inflow plane,
leading to a sheath of marginally relativistic flow.   We note that we
are using a second order scheme for a nearly laminar flow containing
weak structures, and that many studies, starting with that of
\citep{pw94}, have shown that in such cases the numerical viscosity
acts in a way very similar to true viscosity.  Although the magnitude
of the numerical viscosity might not agree with that of the true
viscosity we believe that our current prescription does not build
anomalous behavior into our numerical solution.The simulation was
terminated at dynamical time $\tau_d \equiv$ ta/R = 45.2 when a
quasi-steady state developed on the computational grid.

\vspace{-0.5cm}
\section{Simulation Results}

Figure 1 contains panels showing proper density slice planes transverse
to the jet ($z$) axis. The panels show that the jet is surrounded by a
low density (dark blue) region that grows in transverse extent as
distance from the inlet increases.  Flow is into the page and in these
panels precession and elliptical rotation are counterclockwise and
spiral density (pressure) waves in the external medium can be seen in
the panel at the inlet (top left). Twin higher density filaments are
evident in the panels at z/R = 5 \& 10 as red regions within a yellow
(lower density) envelope.  The filaments rotate by about 225
degrees between these two panels in the clockwise direction.  At larger
distance the twin filaments disappear and only a general elliptical
distortion of the jet along with some displacement of the jet from the
$z$-axis due to helical twisting can be seen in the panels.
\vspace {2.75 in}
\begin{figure}[h!]
\figurenum{1}
\includegraphics{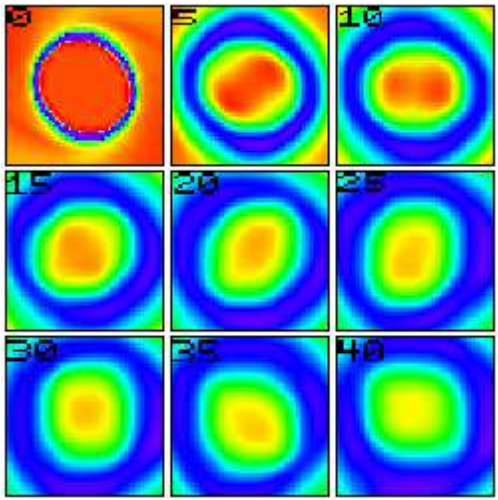}
\caption{\footnotesize \baselineskip 8pt 
Simulation proper density slice planes transverse to the $z$-axis from
the inlet (top left panel) to z/R = 40 (bottom right panel) in $\Delta
z = 5R$ intervals.  Individual panels are of dimension $3.5R \times
3.5R$. In this rendering red to blue is high to low density.
\label{fig1t}}
\end{figure}  

The transverse velocity and density structure of the jet is indicated
more quantitatively by the profiles at $z/R =$ 5, 20 \& 35 shown in
Figure 2.  The approximate edge of the jet (jet spine) is indicated by
the vertical lines in the panels and coincides with rapid change in the
azimuthal velocity component, v$_{\rm y}$.  In general, the azimuthal
velocity changes direction outside the jet spine.  As can be seen from
these transverse profiles, the low density medium outside the jet spine
develops significant outflow that grows in transverse extent and speed
as distance from the inlet increases, i.e., the axial velocity profile
becomes broader relative to the edge of the jet as distance from the
inlet increases.

\newpage
.
\vspace {1.2 in}
\begin{figure}[ht]
\figurenum{2}
\includegraphics{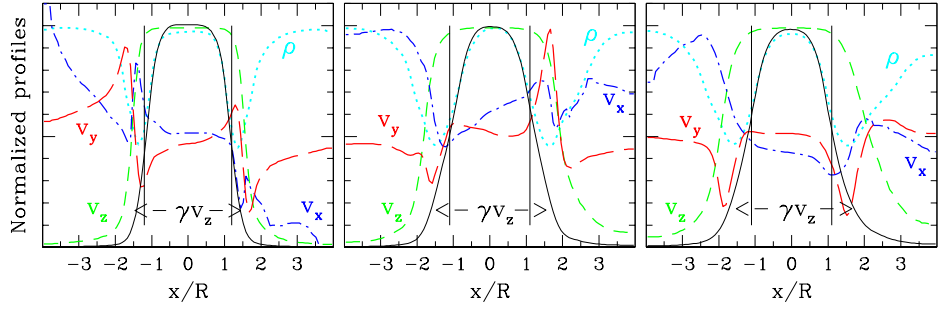}
\caption{\footnotesize \baselineskip 8pt
Transverse normalized profiles across the computational grid along the
$x$-axis are shown at $z/R =$ (left panel) 5, (center panel) 20 \&
(right panel) 35.  Panels include proper density $\rho$ (\Cyan{dotted
line}), jet velocity v$_{\rm z}$ (\Green{shortdash line}), Lorentz
factor times velocity $\gamma$v$_{\rm z}$ (solid line), radial velocity
v$_{\rm x}$ (\Blue{dashdot line}), and azimuthal velocity v$_{\rm y}$
(\Red{longdash line}). The axial velocity component is normalized so
that the maximum is one.  The radial and azimuthal velocity components
are normalized so that the zero point corresponds to the midpoint of
the vertical scale.
\label{fig2t}}
\end{figure}

Plots of the pressure, proper density and velocities along 1-D cuts
parallel to the jet axis are shown in Figure 3.
\vspace {5.0 in}
\begin{figure}[h!]
\figurenum{3}
\includegraphics{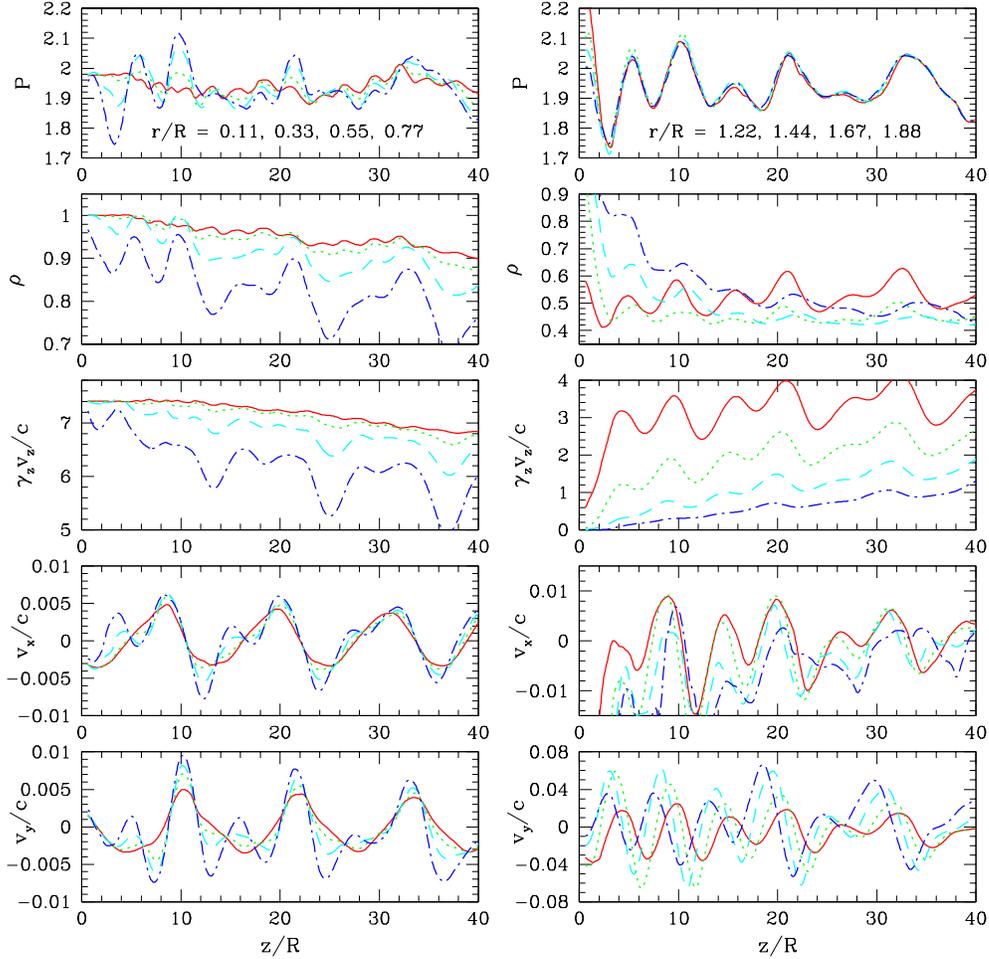}
\caption{\footnotesize \baselineskip 8pt 
Simulation 1-D cuts (left column) at r/R = 0.11 (\Red{solid line}),
0.33 (\Green{dotted line}), 0.55 (\Cyan{dashed line}) \& 0.77
(\Blue{dotdash line}) and (right column) at r/R = 1.22 (\Red{solid
line}), 1.44 (\Green{dotted line}), 1.66 (\Cyan{dashed line}) \& 1.88
(\Blue{dotdash line}).  v$_x$ is a radial velocity and v$_y$ is an
azimuthal velocity.
\label{fig3t}}
\end{figure}

\newpage
\noindent
Inside the jet pressure and density cuts show $\lambda \sim$ 1.5R, 3R,
4R, \& 12R fluctuations.  Transverse velocity component cuts reveal
$\lambda \sim$ 6R \& 12R fluctuations.  The $\lambda \sim$ 4R, 6R \&
12R fluctuations are identifiable with twin density filaments inside
the jet at z/R $\le$ 10, elliptical distortion of the jet cross
section, and with helical twisting of the jet, respectively.  The
$\lambda \sim$~1.5R at z/R~$<$~11 and $\lambda \sim$~3R fluctuations
seen here in the pressure, density and axial velocity are identifiable
with pinching.   There is no significant growth of the initial
perturbation on the computational grid.  Outside the jet the cuts show
that pressure fluctuations are communicated without loss to at least a
jet radius outside the jet spine.  Density cuts show the greatly
reduced density around the jet spine and axial velocity cuts show the
growth in wind speed in the medium outside the jet spine as distance
from the inlet increases.  Radial motions outside the jet spine
increase out to $r/R \sim 1.2$ and then decline, and azimuthal motions
increase out to $r/R \sim 1.7$ and then decline.
  
In the next section we will show that the presence of velocity shear
outside the jet spine, treated as a simple uniform external wind, can
explain much of the behavior seen in the simulation.  If we assume that
the jet spine interacts with the external medium within a layer of
thickness 2R (suggested by a transition to undisturbed ambient at
x/R~$\sim$~3 in the profiles shown in Figure 2 and a theoretically
predicted exponential decline in perturbation amplitude outside the jet
surface), then we may regard the external medium as having a ``wind''
speed $v_{wnd} \geq c/2$ for $z/R > 15$.

\vspace{-0.6cm}
\section{Theoretical Interpretation}

We can analyze flow driven structures, e.g., Hardee (2000), Hardee et
al. (2001), by modeling the jet as a cylinder of radius $R$, having a
uniform density, $\rho _{j}$, and a uniform velocity, $v_{j}$, along
the $z$-axis. The external medium is assumed to have a uniform density,
$\rho _{e}$, and to have a uniform velocity, $v_{e}$, along the
$z$-axis.  The jet is assumed to be in static pressure balance with the
external medium $P_{j}=P_{e}=P_0$. A general approach to analyzing the
time-dependent structures is to linearize the fluid equations along
with an equation of state where the density, velocity, and pressure are
written as $\rho =\rho_0+\rho_1$, ${\bf v}={\bf v}_0+{\bf v}_1$ and
$P=P_0+P_1$, and subscript 1 refers to a perturbation to the
equilibrium quantity. In cylindrical geometry a random perturbation of
$\rho_1$, ${\bf v}_1$ and $P_1$ can be considered to consist of Fourier
components in the source (inlet) frame of the form
\begin{equation}
f_1(r,\phi ,z)=f_1(r)\exp [i(kz\pm n\phi -\omega t)] 
\label{1}
\end{equation}
where the flow is along the z-axis, and $r$ is in the radial direction
with the flow bounded by $r=R$. In cylindrical geometry $k$ is the
longitudinal wavenumber, $n$ is an integer azimuthal wavenumber, for
$n>0$ the wavefronts propagate at an angle to the flow direction, the
angle of the wavevector relative to the flow direction is $\theta =\tan
(n/kR)$, and $+n$ and $-n$ refer to wave propagation in the clockwise
and counterclockwise sense, respectively, when viewed outwards along
the flow direction. In equation (1) $n=$ 0, 1, 2, 3, etc. correspond
to pinching, helical, elliptical, triangular, etc. normal mode
distortions of the jet, respectively. For normal mode $n$ the axial
wavelength associated with a $360^{\circ }$ helical twist of a
wavefront around the jet beam is given by $\lambda _z=n\lambda _n$
where $\lambda _n=2\pi /k$. Propagation and growth or damping of the
Fourier components is described by the dispersion relation
\begin{equation}
\frac{\beta_{j}}{\chi_{j}}\frac{J_{\pm n}'(\beta_{j}R)}{J_{\pm n}
(\beta_{j}R)}=
\frac{\beta_{e}}{\chi_{e}}\frac{H_{\pm n}'
(\beta_{e}R)}{H_{\pm n}(\beta_{e}R)}~~. 
\label{2}
\end{equation}
In the dispersion relation the primes denote derivatives of the Bessel ($J$)
and Hankel ($H$) functions with respect to their arguments, subscripts $j$ and $e$ refer to the jet and external medium, respectively,
$$
\chi_{e}=\gamma_{e}^2\left( \rho_{e}+\frac{\Gamma_{e}}{
\Gamma_{e}-1}\frac{P_0}{c^2}\right) (\omega - kv_{e})^2~~, 
$$
$$
\chi_{j}=\gamma_{j}^2\left( \rho_{j}+\frac{\Gamma_{j}}{\Gamma_{j}-1
}\frac{P_0}{c^2}\right) \left( \omega - kv_{j}\right)^2~~, 
$$
and
$$
\beta_{e}=\gamma_{e}\left[ \frac{(\omega - kv_{e})^2}{a_{e}^2}
-\left( k-\frac{\omega v_{e}}{c^2}\right)^2\right]^{1/2}~~, 
$$
$$
\beta_{j}=\gamma_{j}\left[ \frac{(\omega - kv_{j})^2}{a_{j}^2}
-\left( k-\frac{\omega v_{j}}{c^2}\right)^2\right]^{1/2}~~. 
$$
In the expressions above $\gamma_{j,e} \equiv \left( 1-v_{j,e}^2/c^2\right)^{-1/2}$ is
the unperturbed Lorentz factor, and $a_{j,e}$ is the sound speed given by
\begin{equation}
a_{j,e}\equiv \left\{ \frac{\Gamma_{j,e} P_0}{\rho_{j,e}+[\Gamma_{j,e}/
(\Gamma_{j,e}-1)]P_0/c^2}\right\}^{1/2}~~, 
\label{3}
\end{equation}
where $4/3\leq \Gamma_{j,e} \leq 5/3$ is the adiabatic index, the flow
velocities are measured in the source frame and the densities are
measured in the proper fluid frames. With these definitions the
dispersion relation allows $P>\rho c^2$, sound speeds a large fraction
of lightspeed, and is a simple generalization of previous work (e.g.,
Ferrari et al.\ 1978; Birkinshaw 1991; Hardee et al.\ 1998).

In previous work it has been assumed that the external medium is
stationary relative to the source with $v_e = 0$ and $\gamma_e = 1$ and
typical behavior of solutions to the dispersion relation for
relativistic supersonic flow in a stationary external medium is well
documented in the literature (e.g., Birkinshaw 1991: Hardee et
al.\ 1998; Hardee 2000).  Here in Figure 4 we show solutions to the
dispersion relation for the pinch fundamental (pf), helical (hs) and
elliptical (es) surface, and accompanying first body (b1) mode
distortions to a cylindrical jet with Lorentz factor $\gamma$~=~7.47
(v$_{\rm j}$ = 0.991~c), adiabatic index $\Gamma$~=~13/9 and sound
speed a~=~0.62~c for cases when there is no flow of the external
medium, v$_{\rm e}$~=~0 and $\gamma_e = 1$, and when there is a wind
flow v$_{\rm e}$~=~0.5~c and $\gamma_e \sim 1.15$, of the external
medium relative to the source.  In both calculations the sound speed
and adiabatic index in the external medium are a~=~0.62~c and
$\Gamma$~=~13/9.
\vspace {2.60 in}
\begin{figure}[h]
\figurenum{4}
\includegraphics{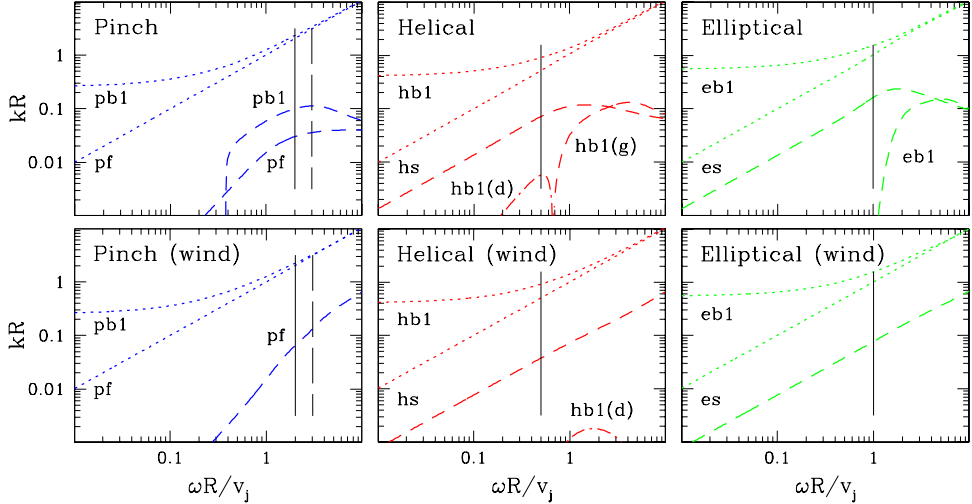}
\caption{\footnotesize \baselineskip 8pt
The real and imaginary parts of the wavenumber are shown by the dotted (real)
and dashed (growing) or dot-dashed (damped) lines, respectively.  The
solid vertical lines indicate the location of the simulation
perturbation frequency or, for the pinch mode, the lowest integer
multiple of the perturbation frequency near to the maximum growth rate
(vertical dashed line) of the pinch body (pb1) mode solution.
\label{fig4t}}
\end{figure}

When there is no wind the helical body (hb1) solution is damped
(dot-dash line) at the precession frequency and growing at higher
frequencies.  With the exception of the pinch fundamental (pf) mode all
solutions have a maximum growth rate at some supersonic ``resonant''
frequency.  With no wind we would expect to see growth of the initial
precessionally driven helical surface (hs) and elliptical surface (es)
perturbations along with body modes (pb1 \& eb1) across the computation
grid.   In the presence of a wind the solutions show that the growth
rates of the surface modes at the precession frequency are suppressed
by about a factor of 2, the body solutions are either purely real or
weakly damped (weak damping of hb1 is shown), the supersonic resonance
has disappeared, and growth of the fundamental pinch mode (pf) has
increased to greater than the growth rate of the surface modes at the
frequencies of interest.
 
The $n=0$ pinch mode fundamental wave solution can be found from the dispersion relation in the low frequency limit $\omega \rightarrow 0$ and $k \rightarrow 0$.  In this limit $\beta_jR \rightarrow 0$,  $\beta_eR \rightarrow 0$ and the dispersion relation becomes
$$
\chi_j \approx -\chi_e(\beta_jR)^2[ln(\beta_eR)-i\pi/2]/2~~,
$$
where we have used 
$[J_{0}^{^{\prime }}(\beta _jR)]/[J_{0}(\beta _jR)] \rightarrow -\beta_jR/2$ 
and
$[H_{0}^{(1)}(\beta _eR)]/[H_{0}^{(1)^{\prime}}(\beta _eR)] \rightarrow
(\beta_eR)[ln(\beta_eR)-i\pi/2]$.  
As $(kR)^2 \rightarrow 0$ faster than
$ln(\beta_eR) \rightarrow -\infty$, the real part of the solution to the dispersion relation remains
nearly unmodified by the presence of a wind around the jet and
\begin{equation}
\frac \omega k\approx v_j~~.
\label{4}
\end{equation}
The imaginary part of the solution is vanishingly small in the low
frequency limit.

The body wave solutions are somewhat modified by the presence of a
wind surrounding a jet and in the limit $\omega \rightarrow
0$ and $k \neq 0$ the dispersion relation becomes
$$
\beta_jR -n\pi/2 -\pi/4 \approx \pm\frac{\pi}{2}\left[1 - C_n\right]~~,
$$
where we have used $J_{n}(z) \approx (2/\pi z)^{1/2}cos(z-n\pi/2-\pi/4)$, $\theta = cos^{-1}\epsilon \approx \pi/2 - \epsilon$,
and $C_n << 1$ is a correction term given by
$$
C_n = \frac{2}{\pi}\frac{\chi _e}{\chi _j}\frac{\beta _j}{\beta _e}\frac{
H_n^{(1)}(\beta _eR)}{H_n^{(1)^{\prime }}(\beta _eR)}\left(\frac{\pi \beta_jR}{2}\right)^{1/2}J_{n}^{^{\prime }}(\beta _jR)~~. 
$$
The solutions are given by
\begin{equation}
kR\approx \frac{(n+2m+1/2)\pi/2-\pi/2(1- C_n)}{\gamma_j \left[M_j^2-1\right]^{1/2}}~~,
\label{5}
\end{equation}
where $m\geq 1$ is an integer. When there is no wind  $\chi _e|_{\omega = 0}=\rho
_ev_e^2k^2 = 0$ and $C_n = 0$.  Previously we found that
unstable body wave solutions exist only when the denominator in
equation (5) is real and that body mode growth rates, with the
exception of the first pinch body mode, are small unless the jet is
sufficiently supersonic in both jet and external medium.

In the low frequency limit $\omega \rightarrow 0$ and $k \rightarrow 0$
where $\beta_jR \rightarrow 0$,  $\beta_eR \rightarrow 0$, the
dispersion relation for the surface modes, $n > 0$, becomes
$$
\chi_j \approx -\chi_e~~,
$$
where we have used 
$[J_{\pm n}^{^{\prime }}(\beta _jR) H_{\pm n}^{(1)}(\beta _eR)]/[
J_{\pm n}(\beta _jR)H_{\pm n}^{(1)^{\prime}}(\beta _eR)] \rightarrow
-(\beta_eR)/(\beta_jR)$.  
Thus, the low frequency
analytical approximation for helical (hs), elliptical (es) and higher
order surface modes is
\begin{equation}
\eqnum{6a}
{\omega \over k} = {v_e+\eta v_j \over 1+\eta } \pm i{\eta ^{1/2} \over
1+\eta }(v_j-v_e)
\label{6a}
\end{equation}
\begin{equation}
\eqnum{6b}
{k \over \omega} = {v_e+\eta v_j \over v_e^2+\eta v_j^2} \mp i{\eta ^{1/2} \over
v_e^2+\eta v_j^2}(v_j-v_e)
\label{6b}
\end{equation}
where
$$
\eta \equiv \left[{\gamma_j \over \gamma_e}\right]^2\left[{a_e \over a_j}\right]^2~,
$$
and growth corresponds to the plus sign in equation (6a) and minus sign
in equation (6b).  The normal mode propagation speed (real part of
eq.[6a]) increases as wind speed increases but the growth rate
(imaginary part of eqs.[6a,b]) decreases as wind speed increases.  The
growth length becomes
\begin{equation}
\eqnum{7}
\ell(\omega)/R \equiv \mid (k_IR)^{-1} \mid = {\eta v_j^2 + v_e^2 \over 
\eta^{1/2}v_j(v_j - v_e)}(\omega R/v_j)^{-1}~.
\label{7}
\end{equation}
It is not surprising that the presence of a wind speeds up wave motion
in the source frame, and reduces the temporal and spatial growth rate
in the source frame through reduction in the velocity shear. Thus,
longer growth lengths in the source frame, equation (7), are not
unexpected.  It is suprising to find that the growth rates and the
growth length can be shown to depend directly on the velocity shear,
$\Delta v = v_{\rm j} -  v_{\rm e}$, independent of the Lorentz
factor.

This analytically predicted effect applies at the precession frequency
in the simulation and explains the reduction in the growth rate of the
surface modes shown in Figure 4.  Note that a high Lorentz factor wind
is not necessary for significant stabilization and is in addition to
stabilization provided by $\eta >>$~1.  If the external medium is
inflowing then $v_e \rightarrow - v_e$ in equations (6a), (6b) and (7),
the growth rate is increased and the growth length decreases in the
source frame.  Structure associated with normal modes can be modeled
theoretically using expressions giving the pressure and velocity
structure (Hardee et al.\ 1998; Hardee 2000; Hardee et al.\ 2001), and
compared to structure seen in the simulation. In Figure 5 we show
suitably normalized 1-D cuts from the simulation along with theoretical
1-D cuts generated using a combination of pinch fundamental (pf),
helical surface (hs), and elliptical surface (es) and first body (eb1)
modes.  The damped helical body solution (hb1) is not included and we
find no evidence for its presence in the simulation.  We have chosen
solutions at the frequencies indicated by the solid lines in Figure 4
and adjusted the amplitudes to correspond as nearly as possible to the
fluctuations observed in the simulation. Note that the amplitudes of
the theoretical and simulation pressure fluctuations are comparable.
Slight differences in location of maxima and minima between theory and
simulation are primarily the result of the use of a constant wavelength
in the theoretical fits whereas the simulation wavelengths change
slightly along the jet as the wind develops.  The simulation cuts in
$\gamma_z$v$_z$ are normalized to show fluctuations
$\Delta(\gamma_z$v$_z$)/c.  The fluctuations along the simulation 1-D
cuts at r/R = 0.55 \& 0.77 in Figure 5 must be multiplied by a factor
of 2 \& 3, respectively.
\vspace {5.0 in}
\begin{figure}[ht]
\figurenum{5}
\includegraphics{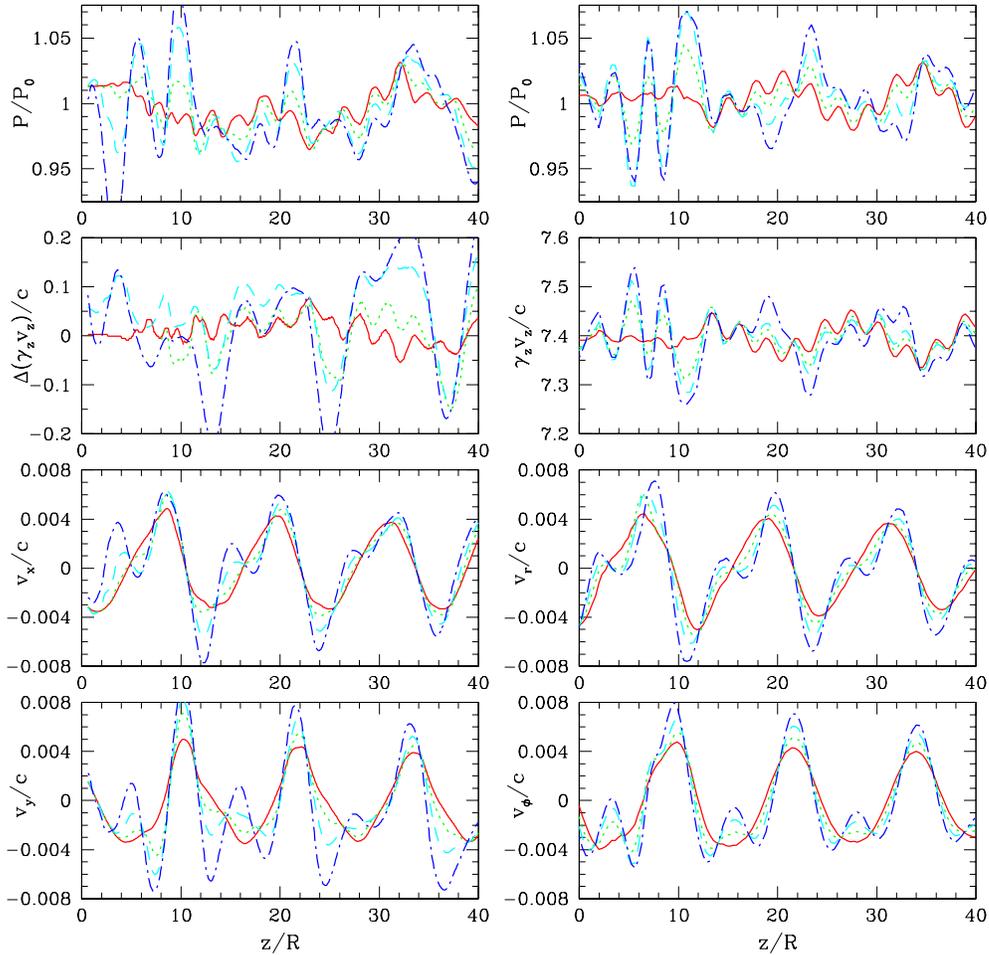}
\caption{\footnotesize \baselineskip 8pt
Normalized simulation 1-D cuts from Figure 3 (left column) and theoretical 1-D cuts corresponding to the simulation 1-D cuts (right column).  1-D cuts are located at r/R = 0.11 (\Red{solid line}),
0.33 (\Green{dotted line}), 0.55 (\Cyan{dashed line}) \& 0.77
(\Blue{dotdash line}). 
\label{fig5t}}
\end{figure}
 
Short wavelength axial velocity and pressure fluctuations ($\lambda
\sim$~3R) are fit by the pinch fundamental mode (pf).  Even shorter
wavelength fluctuations ($\lambda \sim$~1.5R) seen in the simulation at
$z/R < 10$ are consistent with the first pinch body (pb1) mode at the
maximum growth rate.  This solution has not been included in the
theoretical fit.  A long wavelength ($\lambda \sim$ 14R) oscillation in
the pressure corresponds to a standing conical pressure wave pattern at
the relativistic Mach angle.  The elliptical body mode (eb1) at
$\lambda \sim 4R$, evident in the pressure and axial velocity
fluctuations, appears rapidly and then disappears at $z/R > 12$.

Radial and azimuthal velocity fluctuations are almost entirely the
result of helical (hs) and elliptical (es) surface modes and their
amplitudes slowly decline across the computational grid.  In general,
we can match the position of simulation fluctuations to within about a
jet radius in the axial direction.  While there are detail differences
between simulation and theoretical fit, these differences lie primarily
in the axial velocity fluctuations in the outer half of the jet where
the jet Lorentz factor has dropped significantly in the simulation, but
the theory assumes uniform axial velocity and Lorentz factor.  

The fits shown in Figure 5 identify the normal modes conclusively and
also reveal (not shown here) that solutions without a wind fail because
production of the velocity fluctuations observed in the simulation then
requires theoretical pressure fluctuations much higher (factors of 2)
than observed in the simulation.  The disappearance of the elliptical
body mode (eb1) and the pinch body mode (pb1) is a result of the
development of the shear layer at $z/R > 10$.  The growing shear layer,
here modeled theoretically as a uniform external wind, has completely
stabilized these body modes and at least partially stabilized the
helical and elliptical surface modes.

Fitting the fluctuations seen in the numerical simulation involves
computation of theoretical data cubes that can be compared to the
simulation data cubes.  Figure 6 shows pressure slice planes through
the theoretical data cube that can be compared to the simulation proper
density slice planes shown in Figure 1.  In this rendering only the jet
is shown.
%
%
\vspace {2.60 in}
\begin{figure}[h]
\figurenum{6}
\includegraphics{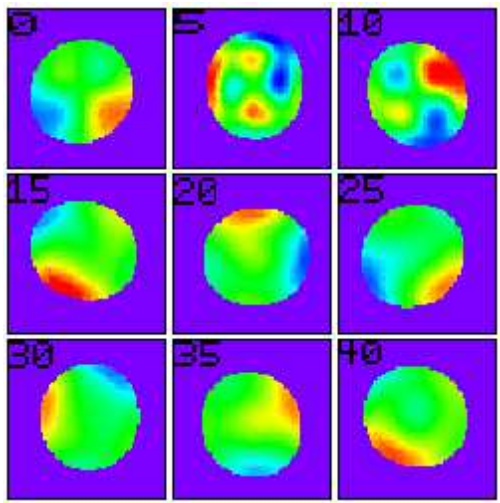}
\caption{\footnotesize \baselineskip 8pt  
Theoretical pressure
slice planes transverse to the $z$-axis from the inlet (top left panel) to z/R = 40 (bottom right panel) in $\Delta z = 5R$ intervals. Panels are of dimension $3.5R \times 3.5R$. In this rendering red to blue is high to low pressure.
\label{fig6t}}
\end{figure}
Twin high pressure filaments are evident in the slice planes at $z/R =
5~ \&~10$.  This is the expected structure for the elliptical body mode
at the precession frequency. Elliptical distortion shows up best in the
slice at  $z/R = 15$ with pressure minima at the ends of the major axis
and pressure maxima at the ends of the minor axis.  The lack of
symmetry reflects interaction with the pressure maximum and minimum
related to helical twisting.  Clockwise rotation of the pressure
maximum associated with helical twisting can be followed in the panels
from $z/R = 20 - 40$.

\vspace{-0.6cm}
\section{Discussion}

In the simulation small perturbations, jet precession and elliptical
cross section distortion, were applied at the inlet to study the
interaction between helical and elliptical jet distortion.  Somewhat
surprisingly, as the jet is predicted to be Kelvin-Helmholtz unstable,
the numerical simulation showed a decline in the amplitude of the
helical and elliptical surface mode perturbations across the
computational grid.   The simulation also revealed a relatively short
wavelength fundamental pinch mode, moving at nearly the jet speed and
with a constant amplitude across the computational grid.  Although
there was internal jet structure associated with pinch and elliptical
body modes just outside the inlet, there was a lack of internal jet
structure in the outer 3/4 of the computational grid.  Previous
simulations at lower Lorentz factor contained internal structure across
the computational grid \citep{hhrg01}.
 
Development of a shear layer of thickness $\Delta r \lesssim$ 2R  at $r
\geq R$  outside the jet spine is responsible for the lack of growth of
the asymmetric modes and for a larger growth rate for the fundamental
pinch mode.   Similar results have been found theoretically and
confirmed by simulation for non-relativistic MHD jets \citep{hr02}.
The effect of a shear layer, modeled as a simple wind, is shown to
reduce the growth of perturbations significantly. Suppression of growth
by a wind is related to reduction in the velocity shear $\Delta v = v_j
- v_e$ and a high Lorentz factor wind is not needed to stabilize a high
Lorentz factor flow. The theory indicates that growth of helical and
elliptical surface mode distortions is reduced by a factor of 2 for a
c/2 wind and would be reduced by a larger factor for the $> c/2$ wind
observed in the simulation at large distances from the inlet. 
Other change in the normal mode solutions within the computed
frequency range, such as the disappearance of a maximum growth rate,
greatly enhanced growth rate of the fundamental pinch mode, and
suppression of the body mode growth rate occurs as the solutions are
very sensitive to small changes in conditions when the velocity shear
is weakly supersonic.
Suppression of the body modes explains the lack of internal jet
structure in the simulation.

Our present results suggest that a suitable shear layer or wind could
partially stabilize a low Mach number relativistic astrophysical jet to
helical and higher order modes of asymmetric jet distortion while
leaving the fundamental pinch mode to grow.  This provides a trigger
for knots moving with the jet speed and with spacing a few times the
jet radius near to the central engine. From this result we conclude
that jets and, by implication, the accretion process onto the central
black hole can be steadier than previously thought, while still
producing rapidly moving knots with quasi-periodic spacing.  We note
that the mechanism found here is quite different from that investigated
by \citet{aetal01}, in which a shock moving with the jet fluid excited
trailing components that could be identified with the first pinch body
mode and were moving much slower than the jet fluid.

Spatial change in the wind speed such as a reduced wind speed at larger
distance from the origin and/or decrease in sound speeds and increase
in the jet Mach number would lead to a change in dynamical behavior.
Rapidly moving knots near to the central engine might trigger and be
replaced by more slowly moving knots associated with rapidly growing
pinch body modes on the supersonic jet.  More rapidly growing
asymmetric structures might also be expected to accompany this change.
Additional rapid growth of asymmetries would accompany jet propagation
through backflow from a high pressure radio lobe as a result of
increase in the velocity shear, $\Delta v = [v_j - (-v_e)]$, and
significant increase in the growth rate (see eqs.[6a,b]) and decrease in
the growth length (eq.[7]).

Knotty jet structure is observed in over half of the extended 3CR
quasars observed by \citet{betal94}.  The jets may be relatively
straight over much of their length, e.g., 3C\,175, or exhibit
significant curvature on the parsec and larger scales, e.g., 3C\,204
[see \citet{betal94}; \citet{hetal99}].  The jet in 3C\.175 (Figure 7)
provides a possible illustration of the effects of external flow on jet
dynamics.
\vspace {3.0 in}
\begin{figure}[h]
\figurenum{7}
\includegraphics{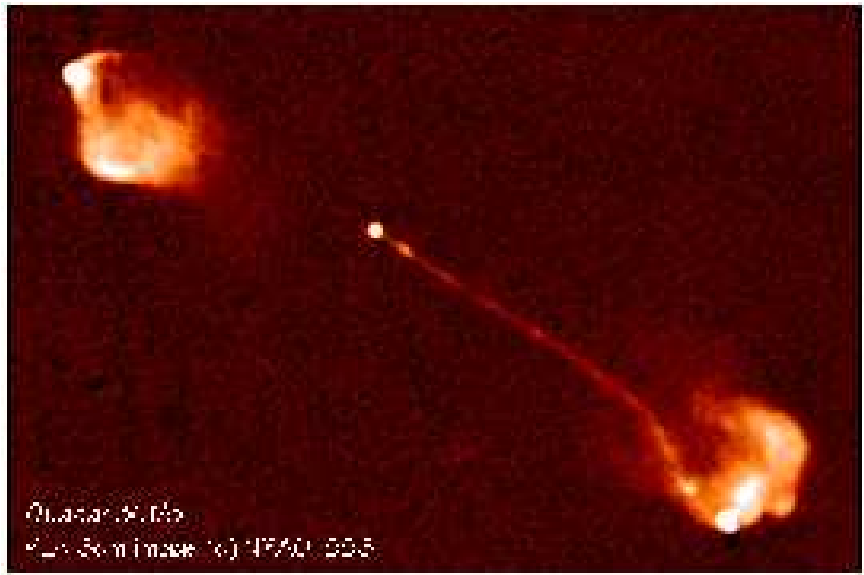}
\caption{\footnotesize \baselineskip 8pt  
Radio image of 3C\,175 at 6~cm (image courtesy of NRAO). 
\label{fig7t}}
\end{figure}  

\noindent
A string of knots is observed along a straight jet that relatively
abruptly bends and terminates in a hot spot and ``U'' shaped region at
the outer edge of the radio lobe. VLBI observations suggest at least
one knot like component beyond the core aligned with the large scale
jet that could be moving superluminally \citep{hetal02}, as would be
expected if knots were associated with triggering of the fundamental
pinch mode in a wind outflow near to the central engine.  At larger
distance the jet is not likely embedded in a rapid outflow and knots
could be slowly moving pinch body modes or simple line-of-sight
crossing effects associated with twisted filaments related to the
elliptical surface and/or elliptical body mode.  There is little
appearance of asymmetric structure before the bend which is coincident
with the trailing edge of the radio lobe in radio intensity images.
The relative abruptness of the bend, although much less abrupt if
deprojected, is suggestive of exponential growth of the helical mode,
possibly triggered by backflow from the high pressure outer edge of the
radio lobe.
 
\acknowledgments

P. Hardee acknowledges support from the National Science Foundation
through grant AST-9802995 to the University of Alabama.

%
%

\vspace{-0.7cm}
\begin{thebibliography}{}
\footnotesize
\baselineskip 8pt
\parskip 0pt

\bibitem [Agudo et al.(2001)] {aetal01} 
Agudo, I., G\'omez, J.L., Mart\'{\i}, J.M., Ib\'a\~nez, J.M., Marscher,
A.P., Alberdi, A., Aloy, M.A., \& Hardee, P.E. 2001, \apj, 549, L183
%
\bibitem[Begelman \& Blandford(1999)]{bb99}
Begelman, M.C., \& Blandford, R.D. 1999, \mnras, 303, L1
%
\bibitem[Birkinshaw(1991)] {b91} 
Birkinshaw, M. 1991, in Beams and Jets in Astrophysics, ed. P.A. Hughes (Cambridge: CUP), 278
%
\bibitem[Bridle et al.(1994)] {betal94} 
Bridle, A.H., Hough, D.H., Lonsdale, C.J., Burns, J.O., \& Laing, R.A. 1994,
\aj, 108, 766
%
\bibitem [Ferrari, Trussoni, \& Zaninetti(1978)]{ftz78} 
Ferrari, A., Trussoni, E., \& Zaninetti, L. 1978, \aap, 64, 43
%
\bibitem[Hardee(2000)]{h00}
Hardee, P.E. 2000, \apj, 533, 176
%
\bibitem[Hardee et al.(2001)]{hhrg01}
Hardee, P.E., Hughes, P.A., Rosen, A., \& Gomez, E. 2001, \apj, 555, 744
%
\bibitem[Hardee et al.(1998)]{hrhd98}
Hardee, P.E., Rosen, A., Hughes, P.A., \& Duncan, G.C. 1998, \apj, 500, 599
%
\bibitem[Hardee \& Rosen(2002)]{hr02}
Hardee, P.E., \& Rosen, A. 2002, \apj, 576, 204 (astro-ph/0205377)
%
\bibitem[Hough et al.(1999)] {hetal99} 
Hough, D.H. et al.\ 1999, \apj, 511, 84
%
\bibitem[Hough et al.(2002)] {hetal02} 
Hough, D.H. et al.\ 2002, \aj, 123, 1258
%
\bibitem[Hughes et al.(2002)]{hmd02}
Hughes, P.A., Miller, M.A., \& Duncan, G.C. 2002, \apj, 572, 713
%
\bibitem[Meier, Koide \& Uchida(2001)]{mku01}
Meier, D.L., Koide, S., \& Uchida, Y. 2001, Science, 291, 84
%
\bibitem[Porter \& Woodward(1994)]{pw94}
Porter, D.H. \& Woodward, P.R. 1994, \apjs, 93, 309
%
\end{thebibliography}
\end{document}